\documentstyle[12pt]{article}
\begin{document}
\hyphenation{me-di-um  as-su-ming pri-mi-ti-ve pe-ri-o-di-ci-ty}
\hyphenation{sti-mu-la-ting con-se-quen-t-ly de-ri-va-ti-ves}
\newcommand{\be}{\begin{equation}}
\newcommand{\ee}{\end{equation}}
\newcommand{\bea}{\begin{eqnarray}}
\newcommand{\eea}{\end{eqnarray}}
\newcommand{\beas}{\begin{eqnarray*}}
\newcommand{\eeas}{\end{eqnarray*}}
\newcommand{\ba}{\begin{array}}
\newcommand{\ea}{\end{array}}
\newcommand{\nn}{\nonumber}
\newcommand{\nonu}{\nonumber}
\newcommand{\bfg}{\begin{figure}}
\newcommand{\efg}{\end{figure}}
\newcommand{\tl}{\tilde}
\newcommand{\rar}{\rightarrow}
\newcommand{\lar}{\leftarrow}
\newcommand{\lrar}{\longrightarrow}
\newcommand{\llar}{\longleftarrow}
\newcommand{\fr}{\frac}
\newcommand{\pa}{\partial}
\newcommand{\mb}{\mbox}
\newcommand{\lft}{\lefteqn}
\newtheorem{th}{Theorem}
\newtheorem{lm}{Lemma}
\newtheorem{cl}{Corollary}
\newtheorem{df}{Definition}
\newtheorem{E}{Example}
\newcommand{\bth}{\begin{th}}
\newcommand{\eth}{\end{th}}
\newcommand{\blm}{\begin{lm}}
\newcommand{\elm}{\end{lm}}
\newcommand{\bcl}{\begin{cl}}
\newcommand{\ecl}{\end{cl}}
\newcommand{\bdf}{\begin{df}}
\newcommand{\edf}{\end{df}}
\newcommand{\brk}{\begin{rm}}
\newcommand{\erk}{\end{rm}}
\newcommand{\hs}{\hspace}
\newcommand{\vs}{\vspace}
\newcommand{\hst}{\hspace*}
\newcommand{\vst}{\vspace*}
\newcommand{\lb}{\label}
\newcommand{\nl}{\newline}
\newcommand{\np}{\newpage}
\newcommand{\om}{\omega}
\newcommand{\Om}{\Omega}
\newcommand{\al}{\alpha}
\newcommand{\bt}{\beta}
\newcommand{\dt}{\delta}
\newcommand{\eps}{\epsilon}
\newcommand{\veps}{\varepsilon}
\newcommand{\ld}{\lambda}
\newcommand{\Ld}{\Lambda}
\newcommand{\gm}{\gamma}
\newcommand{\Gm}{\Gamma}
\newcommand{\sg}{\sigma}
\newcommand{\Sg}{\Sigma}
\newcommand{\bib}{\bibitem}
\newcommand{\ct}{\cite}
\newcommand{\rf}{\ref}
\newcommand{\abschnitt}[1]{\par \noindent {\large {\bf {#1}}} \par}
\newcommand{\subabschnitt}[1]{\par \noindent
                                          {\normalsize {\it {#1}}} \par}
%
\newcommand{\vX}{{\bf X}}
\newcommand{\vnab}{\mbox{\boldmath $\nabla$}}
\newcommand{\nab}{\nabla}
\newcommand{\rcn}{\vr\cdot\vnab}
\newcommand{\rcB}{\vr\cdot\vB}
\newcommand{\rcE}{\vr\cdot\vE}
\newcommand{\LcE}{\vL\cdot\vE}
\newcommand{\LcB}{\vL\cdot\vB}
\newcommand{\bigtr}{\bigtriangleup}
\newcommand{\vr}{{\bf r}}
\newcommand{\vE}{{\bf E}}
\newcommand{\vH}{{\bf H}}
\newcommand{\vF}{{\bf F}}
\newcommand{\vM}{{\bf M}}
\newcommand{\vA}{{\bf A}}
\newcommand{\vD}{{\bf D}}
\newcommand{\vd}{{\bf d}}
\newcommand{\vB}{{\bf B}}
\newcommand{\vP}{{\bf P}}
\newcommand{\vR}{{\bf R}}
\newcommand{\vS}{{\bf S}}
\newcommand{\vL}{{\bf L}}
\newcommand{\vk}{{\bf k}}
\newcommand{\vn}{{\bf n}}
\newcommand{\vJ}{{\bf J}}
\newcommand{\vK}{{\bf K}}
\newcommand{\vV}{{\bf V}}
\newcommand{\vY}{{\bf Y}}
\newcommand{\CMP}[1]{{\em Comm. Math. Phys.}\ {#1}}
\newcommand{\CQG}[1]{{\em Class. Quantum Grav.}\ {#1}}
\newcommand{\EL}[1]{{\em Europhys. Lett.}\ {#1}}
\newcommand{\IJMPA}[1]{{\em Int. J. Mod. Phys.} A\ {#1}}
\newcommand{\IJMPB}[1]{{\em Int. J. Mod. Phys.} B\ {#1}}
\newcommand{\JMP}[1]{{\em J. Math. Phys.}\ {#1}}
\newcommand{\JOP}[1]{{\em J. of Operator Theory}\ {\bf #1}}
\newcommand{\JPL}[1]{{\em J. Phys.} (Paris)\ {\bf #1}}
\newcommand{\JPC}[1]{{\em J. Phys.: Condens. Matter}\ {#1}}
\newcommand{\JPF}[1]{{\em J. Phys. F: Metal Phys.}\ {#1}}
\newcommand{\MPLA}[1]{{\em Mod. Phys. Lett.} A\ {\bf #1}}
\newcommand{\MPLB}[1]{{\em Mod. Phys. Lett.} B\ {\bf #1}}
\newcommand{\NC}[1]{{\em Nuovo Cimento}\ {\bf #1}}
\newcommand{\NP}[1]{{\em Nucl. Phys.}\ {\bf #1}}
\newcommand{\PH}[1]{{\em Physica}\ {\bf #1}}
\newcommand{\PL}[1]{{\em Phys. Lett.}\ {#1}}
\newcommand{\PR}[1]{{\em Phys. Rev.}\ {#1}}
\newcommand{\PRD}[1]{{\em Phys. Rev.}\ {\bf D#1}}
\newcommand{\PRB}[1]{{\em Phys. Rev.} B\ {#1}}
\newcommand{\PRE}[1]{{\em Phys. Rep.}\ {\bf #1}}
\newcommand{\PRL}[1]{{\em Phys. Rev. Lett.}\ {#1}}
\newcommand{\RMP}[1]{{\em Rev. Mod. Phys.}\ {#1}}
\newcommand{\ZPB}[1]{{\em Z. Phys. B}\ {\bf #1}}
\newcommand{\ZPC}[1]{{\em Z. Phys. C}\ {\bf #1}}

\vst{3cm}

\hfill{\small PRA-HEP 93/13}

\vspace*{1cm}

\noindent{\protect\bf PHOTONIC BAND GAP CALCULATIONS : \vst{0.3cm}\\
INWARD AND OUTWARD INTEGRAL EQUATIONS\vst{0.3cm}\\
AND THE KKR METHOD\footnote{To appear in: ``Confined Electrons and Photons :
New Physics and Applications", ed. by E.~Burstein and C. Weisbuch,
Plenum Press, New York (1993).}}

\addvspace{1.3cm}
\noindent\hst{1in}Alexander
Moroz\footnote{Address after September 1, 1993:
Div.  of Theor. Physics, IPN,
Univ. Paris-Sud, F-91 406 Orsay Cedex, France.}

\addvspace{0.5cm}
\noindent\hst{1in}Institute of Physics CAS, Na Slovance 2\\
\noindent\hst{1in}CZ-180 40 Prague 8,  Czech Republic

\thispagestyle{empty}
\addvspace{1.5cm}
\noindent{\bf 1. INTRODUCTION}

\addvspace{0.5cm}
Our main goals have been to calculate photonic bands on various
periodic dielectric lattices, to clarify some controversy regarding
existence of a photonic band gap for fcc lattice of dielectric spheres
\ct{SHI}, and finally treat impurities in a photonic crystal \ct{AK}.
Motivated by the search for a photonic band gap \ct{Y} we have tried to adapt
one of the standard methods of electron band theory
- namely the scalar bulk Kohn-Korringa-Rostocker (KKR) method
\ct{KKR,M} to the form appropriate for photons \ct{AK,AM,AM1}.
We recall that in the case of photons the role of a (periodic) potential
plays $v(\vr)=\eps(\vr)-\eps_o$. Here $\eps(\vr)$ is
the dielectricity of a medium and $\eps_o$ is its host value which is
assumed to be uniform and {\em homogeneous} \ct{Y,AM}.
A dielectric ``{\em atom}" $V_s$ is called a connected
region where $v(\vr)\neq 0$.
The reason of our choice was that the KKR method proceeds
{\em analytically} as far as possible and enables to go beyond
nearly-free photon and plane wave approximations. The latter were
cast into doubts for discontinuous potentials \ct{SHI},
and by a rigorous proof on the existence of finite
number of gaps in two and higher dimensions for the periodic
Schr\"{o}dinger operator \ct{S}.
The KKR starts with an integral equation which is after expansion in a
suitable basis transformed into an {\em algebraic} one.
A band structure then follows from the conditions of solvability
of the algebraic  equation which is vanishing
of a determinant of a matrix (see below)
which determines dispersion relation and eventually photonic
bands. The electronic KKR method is known to lead to a very
compact scheme if the perturbing periodic potential
$v({\bf r})$ is {\em spherically symmetric} within inscribed
spheres and zero  (constant) elsewhere \ct{KKR,M}.
In the case of electrons already $p$-wave approximation gives
agreement within $2\%$ with experiment.

Another distinguished feature of the method is a separation of {\em pure
geometrical} and {\em scattering} properties of a medium.
Geometrical properties are encoded in
{\em geometrical structure constants}
$A_{lm;l'm'}$, $A_{l'm';lm}=A_{lm;l'm'}{}^*$, characteristic for the
lattice under consideration. They are functions of
energy $\sg$ (for photons $\sg=\om\sqrt{\eps_o}$, where $\om$ is frequency)
and the Bloch momentum $\vk$.
Once the structure constants are known then all that is needed
in the case of lattice of identical scatterers  is to
know scattering properties ({\em phase shifts})
of a {\em single} scatterer.

We shall try to elucidate you a derivation of the photonic
KKR along the lines
of \ct{KKR} and from the multiple-scattering theory (MST) \ct{PW}.
The latter is a formal method for calculation of the spectrum in
a (photonic) crystal under the presence of impurities.
In fact it gives a formal solution to the problem of finding
the spectrum in the case of a collection of
{\em arbitrary} distributed scatterers
of {\em different} shape and of {\em different} scattering properties.
In order to simplify our discussion  dielectric medium
is assumed to be {\em magnetically isotropic} and from now on
we shall set  $\mu(\vr) =1$ as well as $c=1$.
On discontinuities $\Sg$
of $\eps(\vr)$ side limits of $\eps(\vr)$ and its derivatives
as well as limiting values of fields and their derivatives
are assumed to be well defined, too.

\addvspace{1cm}
\noindent{\bf 2. INTEGRAL EQUATIONS}

\addvspace{0.5cm}

In course of their derivation we have found that there still persists some
confusion. This originates from the fact
that unlike the scalar case in the case of photons fields
can change {\em discontinuously} across {\em discontinuities} of
electric and magnetic permeabilities and fields are {\em different}
on {\em different} sides of discontinuities.
This is the {\em essential} difference between the Schr\"{o}dinger and
the Maxwell equations\footnote{Another principle difference is that the
Maxwell equations {\em do not} seperate even if $\eps(\vr)$ is a sum
of functions of a single coordinate.}.
Therefore, in the latter case one has to carefully distinguish between
``{\em inward}" and
``{\em outward}" formulations, i.\,e., whether in a given
surface integral inward or outward limits of fields and their derivatives
at the atom boundary are taken \ct{AK,AM,AM1}.
The inward formalism is basically that of  KKR \ct{KKR}
while outward formalism is originally due to Morse \ct{M}.
In particular we have found that the photonic KKR method proposed
in \ct{LWA} does not serve the purpose.
There fields in  integral equations are {\em outward} limits of fields
with respect to the atom boundary while a scalar product
with {\em inward} limits of fields is taken.

The outward integral equation for $\vE(\vr)$ can be written
as follows \ct{AM},
\be
\vE (\vr)= \oint_{\pa V_{s+}} [\vE(\vr ')\,
(d\vS '\cdot\vnab ') G_\sg (\vr,\vr ') -
G_\sg(\vr, \vr ')\,(d\vS '\cdot\vnab ')\vE (\vr ')].
\lb{other}
\ee
Here $G_{\sg}(\vr,\vr')$ is the Green function of the
Helmholtz equation in a region  $\Om$ with suitable boundary
conditions imposed, defined by
\be
(\vnab^2 +\sg^2)G_\sg (\vr ,\vr ') = \delta (\vr -\vr ').
\lb{green}
\ee
 The same equation also holds for $\vB(\vr)$.
Thus, formally both $\vB(\vr)$ and $\vE(\vr)$ satisfy  the same
integral equation (\rf{other}) outside a dielectric atom $V_s$.
What makes the difference between them are matching conditions across a
boundary of the atom.

Corresponding inward integral equation for $\vE(\vr)$ is \ct{AM}
\bea
\lft{\vE(\vr) = \chi_{V_s}(\vr)\,\vE(\vr) +
\oint_{\pa V_{s-}}\left[G_\sg (\vr ,\vr ')\,(d\vS'\cdot\vnab')\vE(\vr ')-
\vE(\vr ')(d\vS'\cdot\vnab')\,G_\sg (\vr ,\vr ')
\right]}\nn\\
&&\mbox{}-\fr{1}{\eps_o}\oint_{\pa V_{s-}}\, [\vnab'G_\sg(\vr ,\vr ')]
v(\vr ')\,(\vE(\vr ')\cdot d{\bf S}')-
\oint_{\pa V_{s-}}G_\sg(\vr ,\vr ')
[\vnab '\cdot\vE(\vr ')]\,d\vS ',\hs{0.5cm}
\lb{elf1}
\eea
$\chi_{V_s}(\vr)$ being the characteristic function of $V_s$ \ct{AM}.

The equivalence of the two expressions can be established either by
taking different limits in integral equations for the vector
potential $\vA(\vr)$ \ct{AM}
or directly by looking at the matching rules
of fields and their derivatives on different sides of $\Sg$.
It can be found in almost any textbook that
\be
\left.(E_n^+ - E_n^-) \right|_\Sg =\fr{\eps_- -\eps_+}{\eps_+} E_n^-
=\fr{v(\vr)}{\eps_o}E_n^-, \hs{0.8cm}E_t^+=E_t^-.
\lb{uno}
\ee
However,  none of the standard textbooks (as far as we know)
tells you how derivatives of fields behave on a discontinuity of
permeabilities. Our results are that normal derivative
$\pa_n \vE_t(\vr)$ of tangential component of $\vE(\vr)$ changes
{\em discontinously} across $\Sg$,
\be
\left.(\pa_n \vE_t^+ -\pa_n \vE_t^-)\right|_\Sg =\left(\fr{1}{\eps_+}-
\fr{1}{\eps_-}\right) \vnab_t D_n =\fr{v}{\eps_o}\vnab_t E_n^-,
\lb{due}
\ee
$\vnab_t D_n(\vr)$ is continuous and
$\pa_n\vE_t^-=\vnab_t E_n^-$.
Rather surprisingly, albeit normal component $E_n(\vr)$ of $\vE(\vr)$ is
discontinuous,
\be
\left. (\pa_nE_n^+ -\pa_nE_n^-)\right|_\Sg =
-\left(\fr{1}{\eps_+^2}(\pa_n\eps)_+ -\fr{1}{\eps_-^2}(\pa_n\eps)_-\right),
\lb{tre}
\ee
and provided $(\pa_n\eps)_+=(\pa_n\eps)_-=0$ the derivative changes
{\em continously} \ct{AM}.
In this manner singularities appearing due to
discontinuities of permeabilities are {\em safely} treated.
These relations explain why the Gauss theorem
which requires at least continuous functions
can be applied to electromagnetic fields despite that fields
change discontinuosly.

\addvspace{1cm}
\noindent{\bf 3. PHOTONIC KKR METHOD}

\addvspace{0.5cm}
%
In the case of photons the KKR method has been only used
within the scalar approximation to the Maxwell equations \ct{JR}.
Following the lines of \ct{KKR} we have formulated a general  variational
principle for the Maxwell equations which holds for an {\em arbitrary
shape} of the basic ``atom'' of a dielectric lattice.
The photonic analogue of the scalar KKR functional is defined
to be
\bea
\lft{\Ld = \om^2\,\int_{V_s} d^3\vr\,v(\vr)\,\vE^*(\vr)\cdot \left\{\vE(\vr)
\mbox{} + \om^2\,
\int_{V_s} G_{\sg}(\vr ,\vr ')\,v(\vr')\,\vE (\vr ')d^3\vr ' +\right.}
\hs{4cm}\nonumber\\
&&\left. \mbox{} + \int_{V_{s+}} \vnab'
G_{\sg}(\vr ,\vr ') [\vnab'\cdot\vE (\vr ')] d^3\vr '
\right\}.
\lb{kkrf}
\eea
The integral over $\vr$ is well defined and does exist as
a {\em bothside} limit.
Provided $\eps(\vr)$ is {\em real} (no absorbtion), and
using well-known hermitian properties of Green's functions,
\(\pa_{\vr}G_\sg (\vr ,\vr ')=-\pa_{\vr '}G_\sg (\vr ,\vr ')\) and
\(G^*_\sg(\vr ,\vr ') =G_\sg(\vr ',\vr)\) one can check that variations
of $\Ld$ with respect to $\vE(\vr)$ or $\vE^*(\vr)$ reproduce correctly the
equation for electric field $\vE(\vr)$ or its complex
conjugate $\vE^*(\vr)$ within $V_s$, respectively.
Rather surprisingly, unless $\eps(\vr)$ is piecewise constant, i.\,e.,
the classical {\em muffin-tin potential}, there is {\em impossible}
(at least for our variational principle) to write
(\rf{kkrf}) in terms of surface integrals only.
Thus, when working with the variational KKR functional it is necesary
to confine ourselves to the case when $\eps(\vr)$ is
nothing but a {\em real muffin-tin potential}.
Then in the case of spherically symmetric dielectric atoms
general formula (\rf{kkrf}) can be simplified further
as follows,
\bea
\lft{\Ld := \lim_{\eps\rightarrow 0}
\oint_{r=r_s-2\eps} dS\, \left[\pa_r \vE^*(\vr) -\vE^*(\vr)\pa_r
\right]\cdot
\left\{\oint_{r'=r_s -\eps}dS'\, \left[\pa_{r '}
\vE(\vr ')-\vE(\vr ')\,\pa_{r '}\right]G_\sg (\vr ,\vr ')
\right.}\hs{4cm}\nn\\
&&\left.\mbox{}-\fr{1}{\eps_o}\oint_{\pa V_{s-}}\,[\vnab ' G_\sg(\vr ,\vr ')]\,
v(\vr ')\,(\vE(\vr ')\cdot d{\bf S}')\right\}.\hs{2.5cm}
\lb{final}
\eea
The final expression (\rf{final}) resembles
the scalar case \ct{KKR}, the only difference being the last term.
The formula (\rf{kkrf}) is called {\em off-shell} while the later
(\rf{final}) is {\em on-shell}.
The reason behind this is that the
latter formula is derived from (\rf{kkrf}) by using field equations
which for a (relativistic) free particle means that it is on
its mass-shell.
Therefore in looking for an extremum of (\rf{kkrf})
one is allowed to use ``{\em arbitrary}" test functions while in the case
of (\rf{final}) one has to ensure that the test functions are
{\em local solutions} to the Maxwell equations. In the latter case
the variational principle selects among the local solutions
that which satisfies prescribed (Bloch) boundary  conditions.

To find the local solutions is rather difficult. Usually only
inward solutions can be found.
{}From now on the further treatment along lines in \ct{KKR}
is straightforward, albeit more involved \ct{AK}.
To proceed  further analytically
$\vE(\vr)$ and $G_\sg (\vr ,\vr ')$ are expanded into
spherical harmonics \ct{J,BL} after which the integral equations
are transformed into matrix equations for expansion
coefficients. If it is done in (\rf{elf1}) one obtains
(in our notation) {\em direct} KKR method while starting
from (\rf{final}) {\em variational} KKR method \ct{AK,AM}.
The ``direct"  photonic KKR method has a wider region of applications :
complex and nonconstant $\eps(\vr)$ within atoms, while
the variational KKR method
has a rather limited range of applications.  However, being variational
it is expected to converge more rapidly within its range of application.

\addvspace{1cm}
\noindent{\bf 4. MULTIPLE-SCATTERING THEORY}

\addvspace{0.5cm}
Standard assumption of MST is that a given medium
can be divided into {\em non-overlapping} spheres $V_n{}'^s$
each of which contains {\em one and only one} scatterer.
MST implicitely assumes that the $t$-matrix ({\em transition factor} \ct{PW})
or phase shifts for a single scatterer are known.
Once they are known
it is convenient to use the {\em outward} integral
equation which is simpler than the inward one.
However, one must not forget that the problem of finding
inward local solutions still remains - they are necessary to
determine just the $t$-matrix!
Following a  standard derivation \ct{PW} one
starts with the $on-shell$ Lippmann-Schwinger equation for $\vE(\vr)$
\ct{AM},
\be
\vE_o(\vr)=
\sum_n\oint_{\pa V_{n+}}dS' [\pa_{r '} G_\sg (\vr,\vr ') -
G_\sg(\vr, \vr ')\,\pa_{r '}]\vE (\vr '),
\lb{other1}
\ee
which holds provided $\vr$ stays inside spheres.
$\vE_o(\vr)$ is  an {\em incident} wave and $n$ labels scatterers (or spheres
containing single scatterers, as you want).
After expanding fields and the Green function into spherical harmonics
one arrives at the
{\em basic photonic MST equations} \ct{AM1},
\be
C^o_{iAL}=\sum_{jA'L'}\left[\delta_{ij}\delta_{LL'}\delta_{AA'}+
i\sum_{A"L"}G^{ij}_{AL,A"L"}t^j_{A"L",A'L'}\right]C_{jA'L'}.
\lb{mst}
\ee
Here indices $i,\,j$ label different scatterers, $G^{ii}\equiv 0$,
and $C^o_{jAL}$ or $C_{jAL}$ are expansion constants of $\vE_o(\vr)$
or $\vE(\vr)$ into spherical harmonics around $j$-th sphere,
respectively.
In contrast to the scalar case the expansion is more involved
due to vectorial nature of fields. This is reason for additional
index $A$ which labels magnetic ($A=M$) and electric ($A=E$)
multipoles \ct{BL}. As a result the order of matrix (\rf{mst})
in a given ($s$-, $p$-,\ldots -wave approximation) is doubled
with regard to the scalar case.
$G^{ij}_{AL,A"L"}$ is the vector (propagator) Green function.
Physically it describes what amount of a particular multipole field
{\em scattered} from the $i$-th site contributes to the particular multipole
field {\em incident} on the $j$-th site.

To write the {\em vector} Green function
$G^{ij}_{AL,A'L'}$ explicitely we have defined quantities
$\bar{C}^\al$  in terms of $3j$-{\em symbols}
$C^\al(l'm'lm)$ \ct{BL},
\bea
\bar{C}^\al(l-1,m+\al)&=&\sqrt{l+1}\,C^\al(l-1,m+\al,l,m),\nn\\
\bar{C}^\al(l+1,m+\al) &=& -\sqrt{l}\,C^\al(l+1,m+\al,l,m),
\eea
and
\be
\bar{T}^\al_{lm+\al}=\fr{1}{\sqrt{l(l+1)}}\,T^\al_{lm+\al},
\ee
where $T^\al_L$ are defined by action of spherical components $\vL^\al$
of orbital angular momentum on spherical harmonics $Y_L$,
$\vL^\al\,Y_L=T^\al_{lm+\al}Y_{lm+\al}$ \ct{AK,AM1}.
Radial functions which are proportional to
$j_l(\sg r_i){}'^s$ (the spherical Bessel functions) are normalized
to unity on the boundary of the $i$-th sphere. This simply
amounts to {\em redefinition} of $C^o_{iML}$ and $C_{iML}$.
We also {\em redefine} $G^{ij}_{EL,EL'}$ by dividing factor
$\sg^2$ to which $\vJ_{EL}{}'^s$ are normalized
(they have different dimension than $\vJ_{ML}{}'^s$ because of
rotation operation applied).
Afterwords the resulting expresions can be written
in the following compact form
\be
G^{ij}_{ML,ML'}=i \sum_{\al=-1}^1
g^{ij}_{l'm'+\al;lm+\al} \bar{T}^\al_{lm+\al}\bar{T}^\al_{l'm'+\al},
\ee
\be
G^{ij}_{ML,EL'}=\sg \sum_{\stackrel{p'=-1}
{\scriptscriptstyle p'\neq 0}}^1 \sum_{\al=-1}^1
g^{ij}_{l'+p',m'+\al,lm+\al}\bar{C}^\al(l'+p',m'+\al)\bar{T}^\al_{lm+\al},
\ee
\be
G^{ij}_{EL,ML'}= -\fr{1}{\sg}\,\sum_{\stackrel{p=-1}
{\scriptscriptstyle p\neq 0}}^1 \sum_{\al=-1}^1
g^{ij}_{l',m'+\al,l+p,m+\al}\bar{C}^\al(l+p,m+\al)\bar{T}^\al_{l'm'+\al},
\ee
\be
G^{ij}_{EL,EL'}=i\sum_{\al=-1}^1\sum_{\stackrel{p=-1}
{\scriptscriptstyle p\neq 0}}^1
\sum_{\stackrel{p'=-1}{\scriptscriptstyle p'\neq 0}}^1
 g^{ij}_{l'+p',m'+\al,l+p,m+\al}
\bar{C}^\al(l+p,m+\al)
\bar{C}^\al(l'+p',m'+\al).
\ee
Note that $G^{ij}_{AL,A'L'}$ {\em is not} Hermitian. This is a consequence
of using the $t$-matrix which also is not Hermitian \ct{KKR,PW}.
For spherically symmetric scatterers the $t$-matrix is {\em diagonal} and for
homogeneous spheres it is explicitely known as the {\em Mie solution}
\ct{RN}.

Provided scatterers are identical and arranged in a {\em periodic} way one
can take Fourier transform with regard to the Bloch momentum $\vk$.
The condition of existence of a solution to (\rf{mst}) for $C^o_{iAL}=0$,
\be
\det\, \left|\delta_{LL'}\delta_{AA'} + i\sum_{A"L"}G_{AL,A"L"}(\vk)
t_{A"L",A'L'}\right| =0,
\ee
then gives the {\em photonic KKR equation}.

We have noted an attempt to derive MST for photons \ct{XZ}.
However one can find obvious difference :
our expression is much more symmetric while in \ct{XZ}
there is no summation over $p,\,p'$ which is necessary here
since electric multipoles of order $l$ have nonzero matrix elements
with $Y_{L'}$ only for $l'=l\pm 1$. Moreover, in our case vector structure
constant are determined via $3j$-symbols and $T^\al_L$ coefficients
in contrast to the Gaunt numbers in \ct{XZ}.
%
%

Presented results use as much as possible existing electron
structure constants and hence allow for a direct numerical
application.  They allow to calculate photonic bands and to
treat impurities as well.
Despite that a photonic band calculation
is more involved due to its vectorial nature there is one
small advantage with regard to the Schr\"{o}dinger equation :
the Maxwell equations are {\em conformally invariant} and therefore
a scale of a dielectric lattice can be chosen arbitrary.
By taking the scale identical to that of electrons
the scalar structure constants can be used without any changes.
For more complete treatment of the above problems together
with numerical results see \ct{AK}.

Last but not least some new phenomenon in a photonic
crystal which has not been discussed yet
and  which may be of some importance : particles propagating
through such medium can emit Tcherenkov radiation at rather small velocity.

\addvspace{1cm}
\noindent{\bf ACKNOWLEDGEMENTS}

\addvspace{0.5cm}
%
Financial support
of the Swiss National Foundation in the early stage of this work
as well as a partial support by the CAS grant No. 11086
are gratefully acknowledged. My thanks is also to the organizers of the
School for preparing very friendly and stimulating enviroment.

%

\end{document}